# Spectral characteristics of side face excited microstructured fibers for photonic integrated circuits formations

I. V. Guryev, D. J. Vazquez, I. A. Sukhoivanov, C. H. Luna, J.M. Esdudillo-Ayala, J. A. Andrade Lucio, R.M. Chavez, M. Trejo-Durán,  E. Alvarado Mendez, R. Rojas-Laguna

Engineering Division, Campus Salamanca, University of Guanajuato
Salamanca-Valle de Santiago Road, Km 3.5+1.8 km, Palo Blanco Community
Salamanca, Guanajuato. C.P. 36885, Mexico

e-mail: iguru@list.ru

**Abstract**: We propose a new method for mass production of the photonic crystal devices on the basis of widely-known and well-developed technology such as micristructured optical fibers. In this paper, we investigate the optical properties of side-excited microstructured fiber and discuss the conditions for utilization such a structure as planar photonic crystal device, namely, the high-quality resonance filter.

**Introduction**

Nowadays, one of the most important challenges concern mass production of the photonic crystal (PhC) devices. At the moment, there are several kinds of the PhC devices available for mass production. Among them there are the artificial opals [1, 2] including inverted opals [3] and the PhC fibers [4]. However, the artificial opals are strictly periodical 3D PhCs with no possibility of the controllable defect introduction and, therefore, cannot be used for the mass production of the PhC devices.

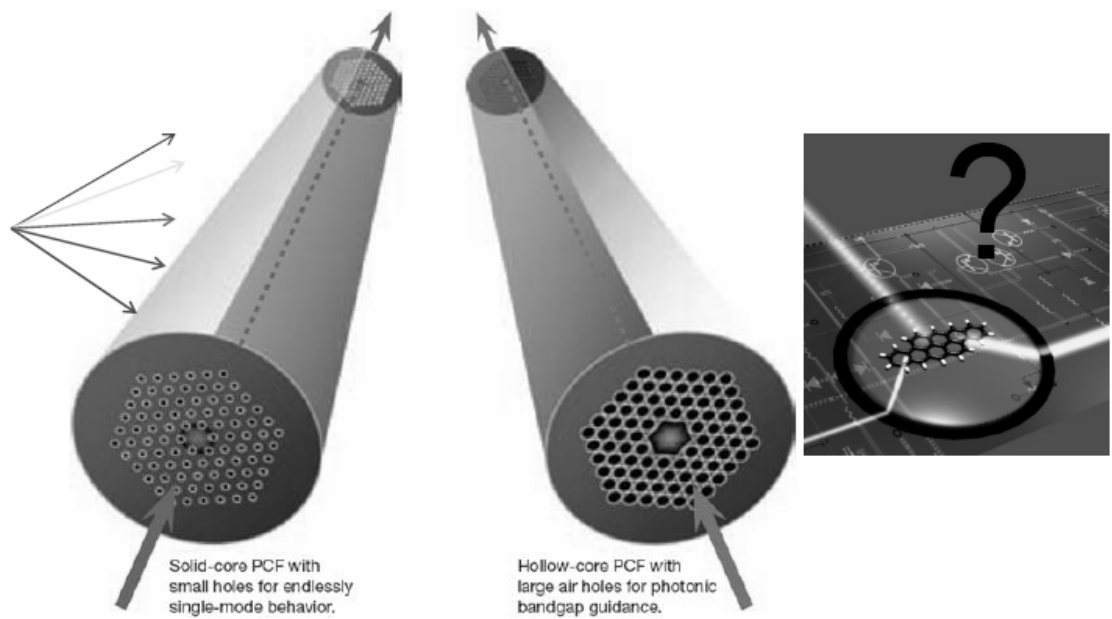

Fig. 1 - The illustration of side-excited PCF used as an element of the photonic integrated circuit

The technology of the PhC fibers though allows production of the PhC fibers with extremely high accuracy and with any profile the developers can imagine. Such an advantage is proved by dozens of years of production of the convenient fibers for telecommunications.

We propose an idea to use the PhC fiber production technology for mass production of the PhC devices. For this reason the fiber may be used in unusual way. Namely, the fiber should be irradiated from the side rather than the face (see fig. 1). In this case the fiber microstructure will behave as a planar PhC with the defect. However, fiber production technology provides the possibility to create fibers with an arbitrary profile. Using such a technology, the profile of the integrated photonic circuit may be easily created. The resulting fiber with complex profile can then be sliced and each piece can be used as a photonic chip after connecting the waveguides to the input and output ports of the scheme.

For this reason, during this work we have been carried out the experimental investigations of the side-face excited fibers and propose theoretical prediction of the fiber behavior which possess properties of the PhC with point defect.

**Experimental investigation of the side-face diffration of the PCF**

The main difference of the PhC from the uniform medium is presence of the diffraction which appears due to the non-linear dispersion characteristic which is always displayed in form of the band structure similar to the one of the solid-state materials (see fig. 2). With this, the radiation wave vector inside the PhC strongly depends on its frequency. This fact leads to differences in diffraction of the radiation observed at different wavelengths.

In the band structure, it is seen that even in case of silica PhC fibers, i.e., low refractive index contrast, we have the band structure which differs from the one in the air which lead to specific properties of such a medium. Particularly, in the PCF the bands splitting is observed between the first and the second bands which leads to partial band gaps appearance and, correspondingly, to the radiation wavelength-sensitivity when propagating across the PCF.

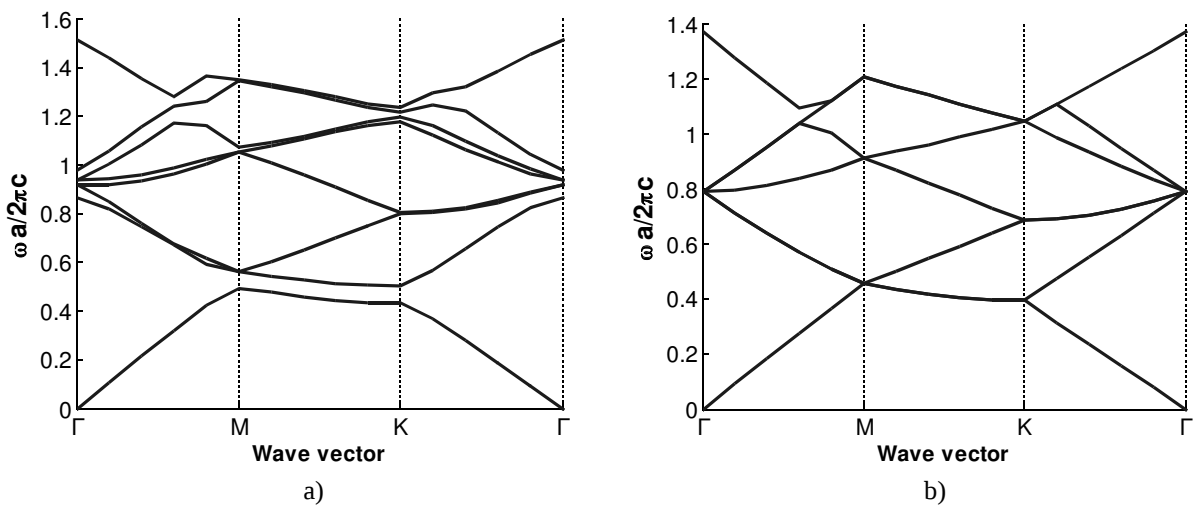

Fig.2 – band structure of the PhC cladding of the side-excited PCF (a) and the one of the bulk silica (b)

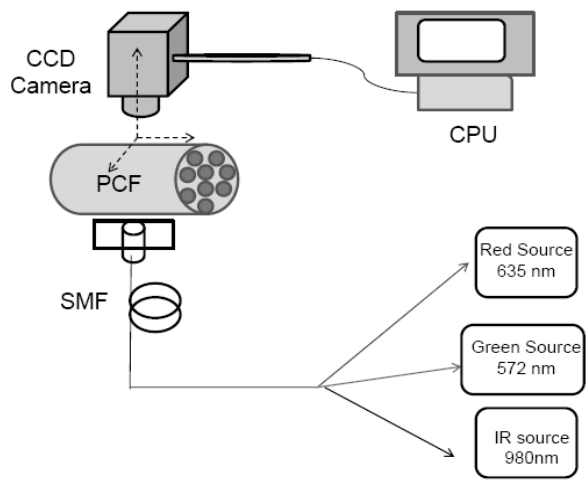

Fig.3 – the scheme of the experimental setup for the diffraction measurement inside the side-excited PCF

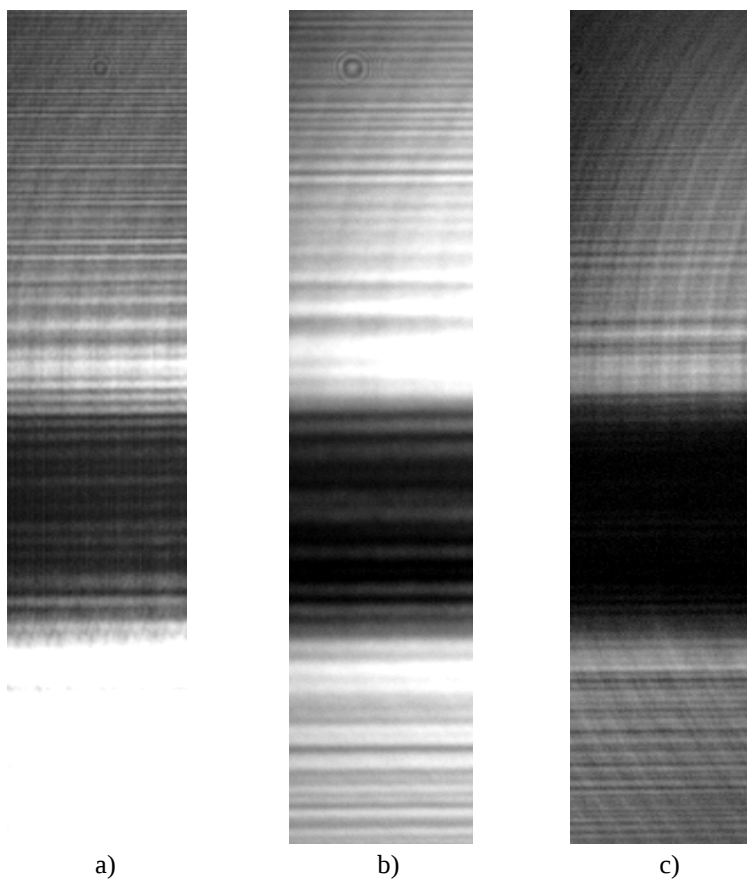

a)          b)          c)

Fig. 4 – the CCD readings for 0.98 microns (a), 0.68 microns (b) and 0.514 microns (c)

In this work we have investigated the radiation diffraction at side-excited PCF. For this reason, we have used lasers at three different wavelength to illuminate the PhC fibers as shown in fig. 3.

Namely, we used wavelength 0.514, 0.63 and 0.98 microns. The diffraction have been read by the CCD camera and then analyzed by means of MATLAB. The results of the diffraction for the LMA20 fiber obtained at the CCD-camera are shown in fig 4.

In the figure 4, light areas correspond to high transmission. We have observed the non-uniform transmission distribution due to the presence of the PCF microstructure in the LMA 20 fiber.

The automatic analysis of the pictures have given the radial intensity distribution for each wavelength shown in the fig. 5. The analysis consisted of reducing the noise, eliminating the non-uniform fiber irradiation and averaging the intensity within the horizontal direction.

In the fig. 5, the central area of the fiber (near r=0) has low transmittance. However, outside the fiber several transmittance peaks are clearly seen. Analyzing the diffraction we can see the maxima intensity shift. With this, in case of green radiation (0.514 microns) the diffraction peaks appear closer to the fiber center than in case of red (0.68 microns) which in turn is closer to the center than infra-red (0.98 microns). Such difference appears due to the non-linear dispersion characteristics as have been predicted previously in this paper.

Although the radiation diffracts in different ways at different wavelengths, this condition is not enough to create fully-functional PhC integrated circuit. The main condition it presence of the photonic band gap which allows to localize the radiation inside the defects and waveguides. However, the band structure computation for such kind of fibers does now show the existance of the band gap (see fig. 2). This situation is typical for low refractive-index contrast PhCs.

**Theoretical models for the PCF made of amorphous silicon**

As we've shown in previous section, the radiation cannot be localized with the PhC fiber made of silica. However, we predicted that the fiber with similar profile is able to localize the radiation inside the defect of the PhC if it is made of amorphous silicon which refractive index is about 4. To prove the PhC properties we have been computed the band structure by means of plane waves expansion method which "master equations" for TE and TM polarizations are following [5]:

$$-\sum_{\vec{G}'} \chi\left(\vec{G}-\vec{G}'\right)\left(\vec{k}+\vec{G}'\right) \times \left\{\left(\vec{k}+\vec{G}'\right) \times \vec{E}_{\vec{k}n}\left(\vec{G}'\right)\right\} = \frac{\omega_{\vec{k}n}^{(E)2}}{c^2}\vec{E}_{\vec{k}n}\left(\vec{G}\right),$$

$$-\sum_{\vec{G}'} \chi\left(\vec{G}-\vec{G}'\right)\left(\vec{k}+\vec{G}\right) \times \left\{\left(\vec{k}+\vec{G}'\right) \times \vec{H}_{\vec{k}n}\left(\vec{G}'\right)\right\} = \frac{\omega_{\vec{k}n}^{(H)2}}{c^2}\vec{H}_{\vec{k}n}\left(\vec{G}\right),$$

(1)

where $\chi\left(\vec{G}-\vec{G}'\right)$ is a Fourier expansion coefficient, $\vec{k}$ is a wave vector within the first Brillouin zone, $\vec{G}$ and $\vec{G}'$ are the reciprocal lattice vectors.

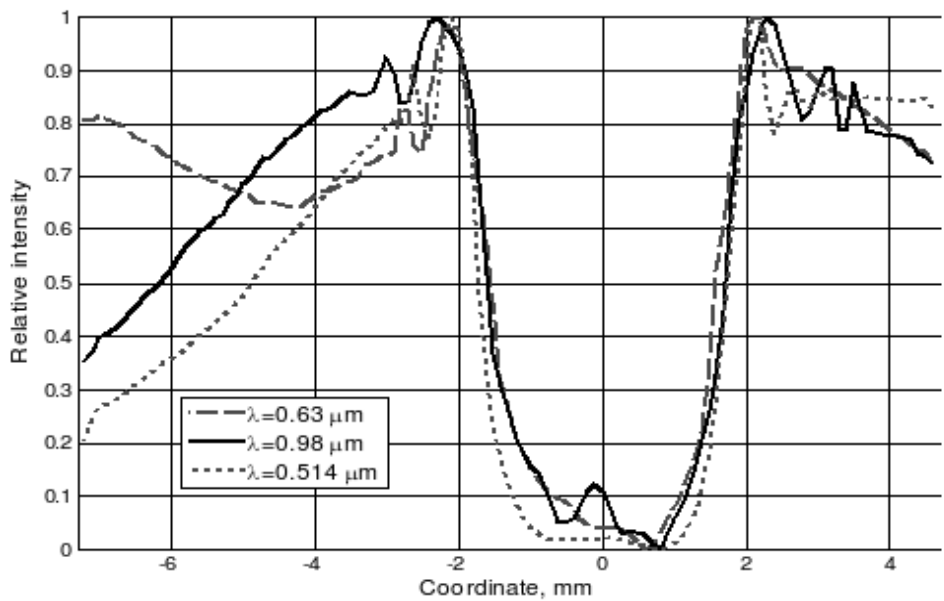

Fig. 5 - results of the automatic analysis of the CCD images obtained at different wavelengths

Solving these equations for the exact values of the wave vectors we obtain the set of the frequencies which give the band structure of the considered PhC without defect and with point defect in the center. The characteristics are shown in fig. 6.

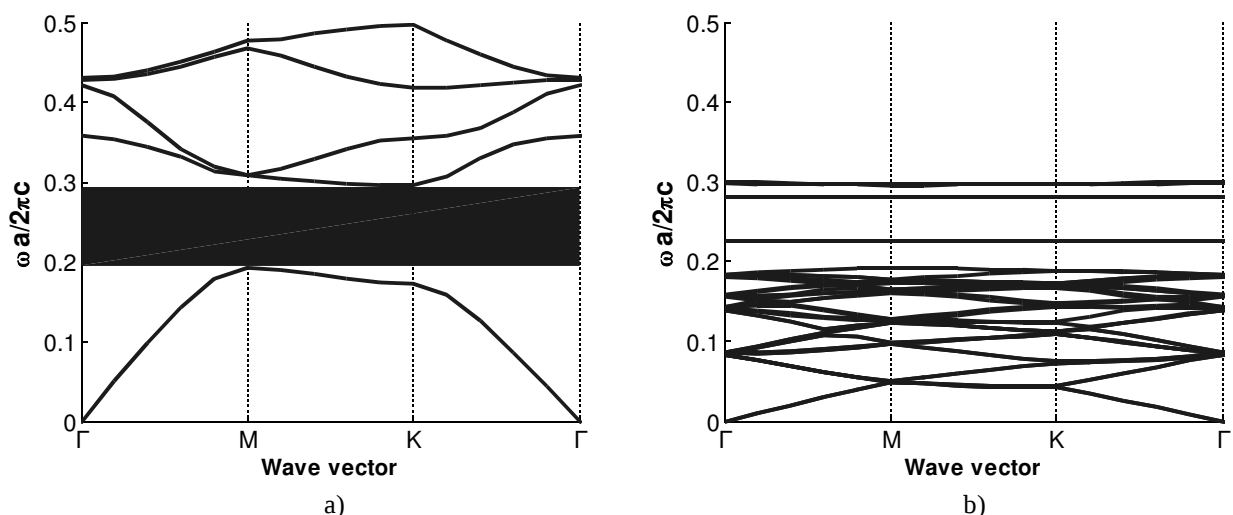

Fig. 6 – the band structures of the bulk PhC (a) and the PhC (b) with defect made of the amorphous silicon

In the Fig. 6a the bandgap appears between approximately 0.2 and 0.3 of relative frequency. Due to this two defect eigen-states appear inside the bandgap when we create the point defect by removing a single hole from the PhC. Such a defects correspond to the transmission maxima.

However the scale of the LMA20 fiber is too large, namely, it's holes diameter is 8 microns. Particularly, the defect eigen-states observed in the band structure correspond to approximately 70 and 57 microns. In order to avoid this kind of problems we also propose to scale the fiber which is allowed by the technological process. As a result, in case of the PhC period about 0.35 we get the eigenstates at the wavelengths 1.25 and 1.55 microns.

The proof of this suggestion was carried out by means of the finite difference time domain (FDTD) method. In our case, the structure is assumed to be uniform within the direction along the fiber and, therefore, the 2D model can be used. The recurrent formula in case of 2D PhC and TM polarization are following [6]:

$$H_{x(i,j)}^{n+1/2} = H_{x(i,j)}^{n-1/2} - \frac{c\Delta t}{\mu \Delta y}\left(E_{z(i,j)}^{n} - E_{z(i,j-1)}^{n}\right),$$

$$H_{y(i,j)}^{n+1/2} = H_{y(i,j)}^{n-1/2} + \frac{c\Delta t}{\mu \Delta x}\left(E_{z(i,j)}^{n} - E_{z(i-1,j)}^{n}\right),$$

$$E_{z(i,j)}^{n+1} = E_{z(i,j)}^{n} + \frac{c\Delta t}{\varepsilon \Delta x}\left(H_{y(i+1/2,j)}^{n+1/2} - H_{y(i-1/2,j)}^{n+1/2}\right) -$$

$$- \frac{c\Delta t}{\varepsilon \Delta y}\left(H_{x(i,j+1/2)}^{n+1/2} - H_{x(i,j-1/2)}^{n+1/2}\right). \quad (2)$$

These equations connect the electric and magnetic field components intensity at different points of the space and time. Solving them recursively, we obtained the time-dependent field distribution within the PhC. The transmittance spectrum investigated PCF have been obtained by the Fourier-transform of the temporal response of the structure to the Gaussian-shaped pulse at the wavelength 1.55 microns (see fig. 7).

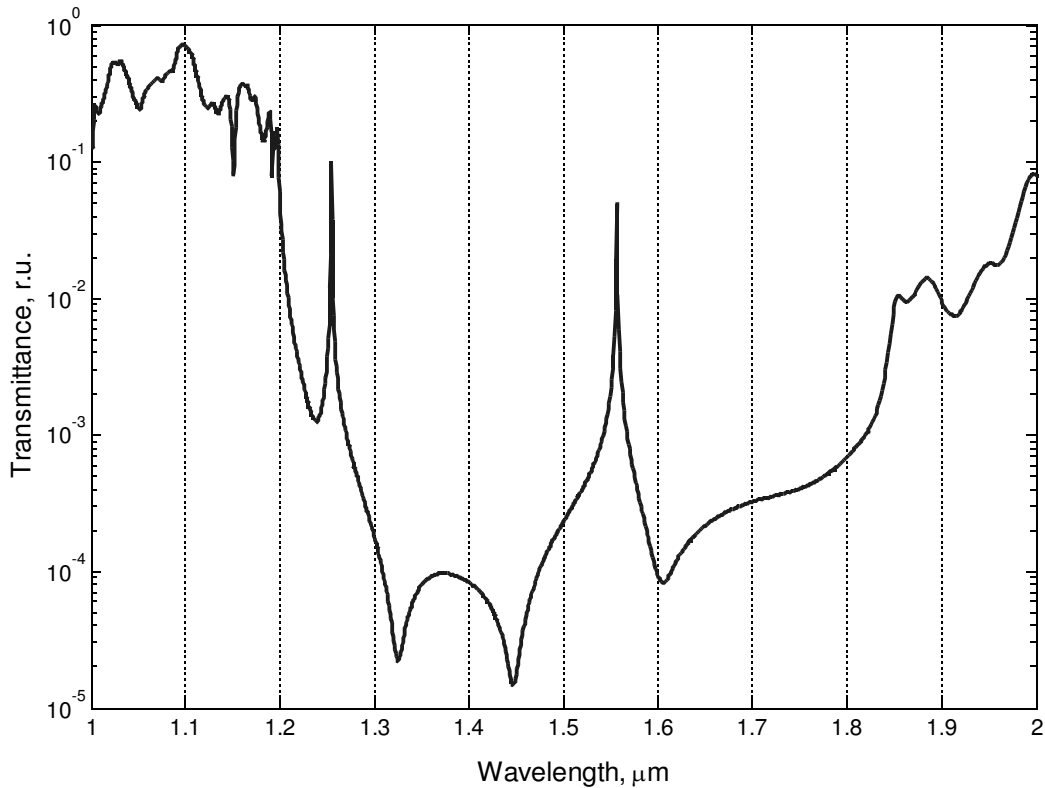

Fig. 7 – the transmittance spectrum of the side-excited PCF made of silicon and with holes scaled down to the sub-wavelength size

**Conclusion**

In the work we have been carried out several experimental and theoretical investigations which have shown the ability to use the side-excited fiber-like structures as the elements of the integrated PhC circuits. There was carried out the parameters fitting on the investigated fiber which allow to use the side-excited PCF as a high-quality PhC filter with transmittance peaks at 1.25 and 1.55 microns.

This idea provides mass production of the photinic integrated circuits of an arbitrary configuration which was previously impossible.

However, we have also shown that achievement of the desired parameters which allows mass production of the PhC elements is impossible without the creation of the PCF with essentially larger refractive index contrast than it is provided by regular PCF. Moreover, the numerical methods should be used for geometric parameters fitting such as holes radius and the pitch.